\documentclass[prl,twocolumn,showpacs]{revtex4}
\usepackage{graphicx,epsf}
\usepackage{bm, bbm}      

\newcommand{\beq}{\begin{equation}}
\newcommand{\beqn}{\begin{eqnarray}}
\newcommand{\eeq}{\end{equation}}
\newcommand{\eeqn}{\end{eqnarray}}
\newcommand{\nn}{\nonumber}

\newcommand{\da}{\downarrow}
\newcommand{\ua}{\uparrow}

\newcommand{\SU}{{\rm SU}}

\usepackage{color}

\begin{document}

\def\tende#1{\,\vtop{\ialign{##\crcr\rightarrowfill\crcr
\noalign{\kern-1pt\nointerlineskip}
\hskip3.pt${\scriptstyle #1}$\hskip3.pt\crcr}}\,}

\title{Atypical Fractional Quantum Hall Effect in Graphene at Filling Factor $\nu_G=1/3$}
\author{Z. Papi\'c$^{1,2,3}$, M. O. Goerbig$^{1}$, and N. Regnault$^{2}$}

\affiliation{
$^1$Laboratoire de Physique des Solides, CNRS UMR 8502, Univ. Paris-Sud, F-91405 Orsay cedex, France\\
$^2$Laboratoire Pierre Aigrain, D\'epartement de Physique, ENS, CNRS, 24 Rue Lhomond, F-75005 Paris, France\\
$^3$Institute of Physics, University of Belgrade, P. O. Box 68, 11\,000 Belgrade, Serbia}

\begin{abstract}

We study, with the help of exact diagonalization calculations, a four-component
trial wave function that may be relevant for
the recently observed graphene fractional quantum Hall state at a filling factor $\nu_G=1/3$.
Although it is adiabatically connected to a 1/3 Laughlin state in the upper spin branch, with
SU(2) valley-isospin ferromagnetic ordering and a completely filled lower spin branch, it reveals physical properties 
beyond such a state that is the natural ground state for a large Zeeman effect. Most saliently, it possesses 
at experimentally relevant values of the Zeeman gap low-energy spin-flip excitations that may be unveiled in 
inelastic light-scattering experiments.

\end{abstract}
\maketitle

The recent observation of the fractional quantum Hall effect (FQHE) in graphene \cite{du09,bolotin09} has proven the 
relevance of Coulomb interactions in this novel two-dimensional (2D) electron system, in agreement with theoretical expectations
\cite{peres06,GMD,AC06,YDSMD06,khvesh07,toke07}. 
The most pronounced state is the one found when the ratio $\nu_G=n_{el}/n_B$
between the electronic density $n_{el}$ and that of the flux quanta $n_{B}=eB/h$ is $\nu_G=1/3$. Although this state is
reminiscent, at first sight, of the prominent 1/3 state observed in semiconductor heterostructures \cite{TSG}, which is described 
to great accuracy by the
Laughlin state \cite{laughlin}, several questions arise when taking fully into account
the four-component strucure of graphene, due to its four-fold spin-valley degeneracy. Whereas first numerical approaches
\cite{AC06} considered the physical spin to be frozen by the Zeeman effect and concentrated on the valley-isospin degree
of freedom in a two-component system, a four-component approach \cite{toke07} 
seems to be more appropriate in view of the rather small
energy scale associated with the Zeeman effect $\Delta_Z$, when compared to the leading energy scale of the Coulomb interaction,
$e^2/\epsilon l_B$ at the magnetic length $l_B=\sqrt{\hbar c/eB}$. Indeed 
for a $g$-factor of 2 \cite{zhang2}, one obtains $\Delta_Z/(e^2/\epsilon l_B)\sim 0.002 \sqrt{B{\rm [T]}}\times
\epsilon$, where $\epsilon$ is the relative dielectric constant. 

A further complication arises in graphene, as compared to the 2D electron gas in semiconductor heterostructures, when one considers 
the definition of the filling factor $\nu_G$, which is proportional to the carrier density $n_{el}$. In graphene, the carrier density
vanishes at the Dirac point, where the spectrum is particle-hole symmetric. In the presence of a magnetic field, a 
four-fold degenerate zero-energy Landau level (LL) is formed that happens to be half-filled when $n_{el}=0$ and thus $\nu_G=0$ --
the situation at $\nu_G=0$ is therefore more reminiscent of a filling factor of $\nu=2$ in a usual four-component quantum Hall system
\cite{toke07}, and the observed FQHE at $\nu_G=1/3$ corresponds to a situation where two of the four spin-valley 
subbranches are completely filled and a third one 1/3-filled ($\nu=2+1/3$). As a consequence the observed FQHE is not a simple
Laughlin state with an SU(4)-spin-valley ferromagnetic ordering, which would arise at $\nu_G=-2+1/3$ (or by particle-hole symmetry,
at $\nu_G=2-1/3$) \cite{YDSMD06,toke07}. A natural candidate for large values of the Zeeman gap would then be a
valley-SU(2)-ferromagnetic Laughlin state $\Psi_{\da,L}^{v-\SU(2)}$ in the spin-$\da$ branch
of the zero-energy LL similar to the usual 1/3 physics. In this scenario, both states $K$ and $K'$ are completely filled in 
the spin-$\ua$ branch. The small relative value of the Zeeman gap, however, casts doubts on such a scenario of
complete spin polarization induced by an external field, without considering a cooperative effect mediated by the
Coulomb interaction.


Here, we analyse the system in the zero-energy LL with the help of 
exact-diagonalization (ED) calculations for relativistic electrons
in the spherical geometry with SU(4) symmetry that
interact via the Coulomb interaction \cite{GMD,AC06,toke07}. We show
that already for a very small Zeeman effect, one may obtain a FQHE at $\nu_G=1/3$ in graphene. This state may be described
in terms of a four-component Halperin wave function $\Psi_{2+1/3}^{\SU(4)}$
which is adiabatically connected to the valley-SU(2)-ferromagnetic state in the upper
spin branch. The latter 
is the natural ground state for a large Zeeman splitting. Most saliently, in spite of this adiabatic connection, the 
low-energy excitations in an intermediate range of the Zeeman splitting are different from those of the simple SU(2) Laughlin state.
In addition to the charge-density-wave mode with its characteristic magneto-roton minimum and the valley-isospin wave, which is
the Goldstone mode associated with the spontaneous valley-isospin breaking in the spin-$\da$ branch, we find a low-energy spin-flip
mode with a gap that depends linearly on the Zeeman coupling. These modes may be experimentally accessible in inelastic light-scattering
measurements that have revealed similar modes in conventional quantum Hall systems in GaAs heterostructures 
\cite{eriksson,perspectives}. That electrons in graphene reside at the sample surface makes this novel 2D electron system even
better adapted to optical measurements than the latter. 

In order to describe the FQHE state at $\nu_G=1/3$, which corresponds to a filling factor of $\nu=2+1/3$ when counted from the bottom 
of the central $n=0$ LL, we investigate the trial four-component wave function  
\beqn\label{wave1}
\nn
\Psi_{2+1/3}^{\SU(4)}&=&\prod_{\xi=K,K'}\prod_{i<j}\left(z_i^{\da,\xi}-z_j^{\da,\xi}\right)^3
\prod_{i,j}\left(z_i^{\da,K}-z_j^{\da,K'}\right)^3\\
&&\times\prod_{\xi=K,K'}\prod_{i<j}\left(z_i^{\ua,\xi}-z_j^{\ua,\xi}\right),
\eeqn
where $z_j^{\sigma,\xi}$ 
denotes the complex coordinate of the $j$-th electron in the spin-valley subbranch 
$\sigma,\xi$ ($\sigma=\ua$ or $\da$ and $\xi=K$ or $K'$). 
We have omitted 
an ubiqitous Gaussian factor in the expression. 
Notice that, in the absence of a symmetry-breaking field, the wave function (\ref{wave1})
is not a good trial state because the Coulomb interaction potential respects the SU(4)-spin-valley symmetry \cite{GMD,AF}, 
whereas the wave function (\ref{wave1}) is not an eigenstate of the SU(4)-Casimir operators. 

This is indeed
corroborated by our ED calculations for $N=17$ particles with $N_B=6$ flux quanta threading the sphere, the relation between
$N$ and $N_B$ being $N_B=(3/7)N - 9/7$ for the state (\ref{wave1}) \cite{DGRG}, which yields the required
filling $\nu=7/3$ in the thermodynamic limit. The ground state is then found in spin sectors different from that,
$2S_z=11$, expected for the state (\ref{wave1}). 
A simple manner to stabilize the state (\ref{wave1}) is to use appropriate pseudo-potentials \cite{haldane} that break the
SU(4)-spin-valley symmetry in the interaction potential. However, surprisingly, this trial state becomes the ground state also
when the SU(4) symmetry is broken by an external field -- e.g. a very small value of the Zeeman effect turns out to be sufficient,
$\Delta_Z^1\simeq 0.01 e^2/\epsilon l_B$, which is a tiny fraction 
of the leading Coulomb interaction energy scale $e^2/\epsilon l_B\sim 625 (\sqrt{B{\rm [T]}}/\epsilon)$ K. 
In a typical experimental situation, the
1/3 state has been found at a magnetic field of roughly $B\sim 9 ... 12$ T \cite{du09,bolotin09}, which corresponds to a ratio of 
$\Delta_Z/(e^2/\epsilon l_B)\sim 0.006...0.008\times\epsilon$ if one uses $g\sim 2$ \cite{zhang2}. This value is slightly smaller than
our theoretical estimate if one considers the smallest possible value of the dielectric
constant ($\epsilon=1$ for free-standing graphene). However, virtual interband excitations lower the dielectric constant
that becomes $\epsilon_{\infty}\simeq 4$ for free-standing graphene in the large-wave-vector limit \cite{GGV}, also in a strong magnetic
field \cite{RFG}, and the precise value of the dielectric constant in graphene remains an open issue.
Notice that it is even still under debate whether the Zeeman splitting is indeed the 
dominant SU(4)-symmetry-breaking perturbation or whether the valley splitting is more relevant. Our theory and the conclusions 
of this paper, however, remain valid also in the latter case if one interchanges the role of spin and valley isospin and if one
replaces $\Delta_Z$ by a ``valley Zeeman effect'' \cite{goerbigRev}.

\begin{figure}
\centering
\includegraphics[width=5.5cm,angle=-90]{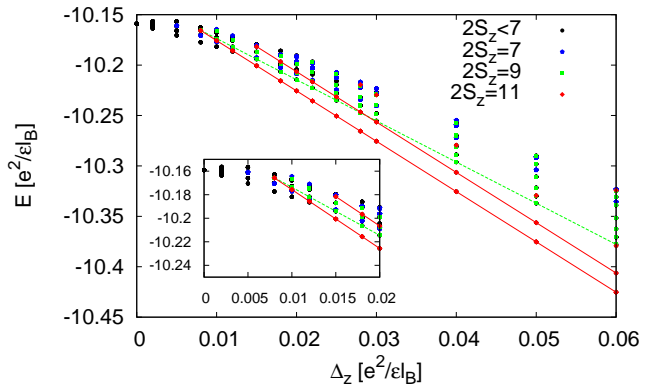}
\caption{\footnotesize{(Color online) Energy spectrum for $N=17$ electrons at a filling factor $\nu_G=1/3$ ($\nu=2+1/3$), as a function
of $\Delta_Z$, obtained from ED calculations of the Coulomb interaction on the sphere ($N_B=6$) with implemented SU(4) symmetry. 
Above $\Delta_Z^1\simeq 0.01 e^2/\epsilon l_B$, the ground state is found in the maximally spin-polarized sector ($2S_z=11$, red diamonds). The inset shows a zoom on the region for small values of $\Delta_Z$.
}}
\label{figSU4}
\end{figure}

\begin{figure}
\centering
\includegraphics[width=6.5cm,angle=0]{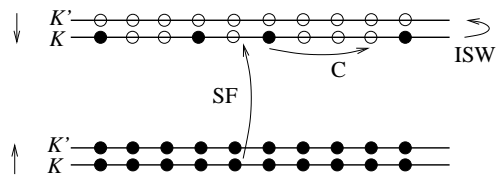}
\caption{\footnotesize{Classification of the excitations of the $\Psi_{1/3}^{\SU(4)}$ state. The excitations of a one-component
Laughlin state are found in the same spin-valley sector (C), whereas the Goldstone mode due to the broken SU(2) valley-isospin 
symmetry in the spin-$\da$ branch is an insospin-wave mode (ISW). In addition to these conventional modes, the four-component
state (\ref{wave1}) possesses a spin-flip (SF) mode.}}
\label{fig01}
\end{figure}

The energy spectrum, which we have obtained in ED calculations, is shown in Fig. \ref{figSU4}
as a function of the Zeeman gap $\Delta_Z$. Above the critical value $\Delta_Z^1$, the ground state is
found in the maximally-polarized spin sector that corresponds to the state (\ref{wave1}), whereas the excited state with the lowest
energy is in the same spin sector, $2S_z=11$, only above a second value $\Delta_Z^2\simeq 0.03 e^2/\epsilon l_B$. For
values of the Zeeman gap $\Delta_Z^1\leq \Delta_Z\leq \Delta_Z^2$, the excited state with lowest energy is found in the 
spin sector $2S_Z=9$. Above $\Delta_Z^2$, however, the energy 
cost of this spin-flip excitation (SF, see Fig. \ref{fig01})
is larger than the lowest-lying excitation in the fully polarized sector $2S_z=11$ (C in Fig. \ref{fig01}).

These results suggest that the state (\ref{wave1}) may have physical properties beyond the simple 1/3-Laughlin state in the spin-$\da$
branch, in the form of coherent spin-flip excitations in an intermediate range of Zeeman gaps. In order to test this 
scenario in more detail, we have investigated the two-component wave function
\beq\label{wave2}
\Psi_{1+1/3}^{\SU(2)}=\prod_{i<j}\left(z_i^{\da}-z_j^{\da}\right)^3
\prod_{i<j}\left(z_i^{\ua}-z_j^{\ua}\right),
\eeq
which would be a candidate in a two-component quantum Hall system, such as a conventional 2D electron gas in a GaAs heterostructure,
at a filling factor $\nu=1+1/3$. It is insofar related to the four-component wave function (\ref{wave1}) as it describes the same
physical situation if the valley-isospin degree of freedom for spin-$\da$ electrons is neglected. The novel wave function (\ref{wave2})
therefore does not reveal any valley-isospin-wave mode (ISW, see Fig. \ref{fig01}) that is the Goldstone mode of the spontaneously 
broken valley-SU(2) symmetry in the spin-$\da$ branch and that may eventually become gapped if one takes into account a possible 
valley splitting. In contrast to its four-component analogue (\ref{wave1}), the two-component 
wave function (\ref{wave2}) allows for a more detailed study of different system sizes in ED with an implemented SU(2) symmetry. Indeed,
our ED calculations with an implemented SU(4) symmetry allowed only for one single system size ($N=17$ particles,
only $N_{\da}=3$ populate the upper spin branch), in which case the
subspace with maximal spin polarization ($2S_z=11$) is of dimension one such that the overlap with the wave function (\ref{wave1}) is
trivially 1. 

\begin{figure}
\centering
\includegraphics[width=7.5cm,angle=0]{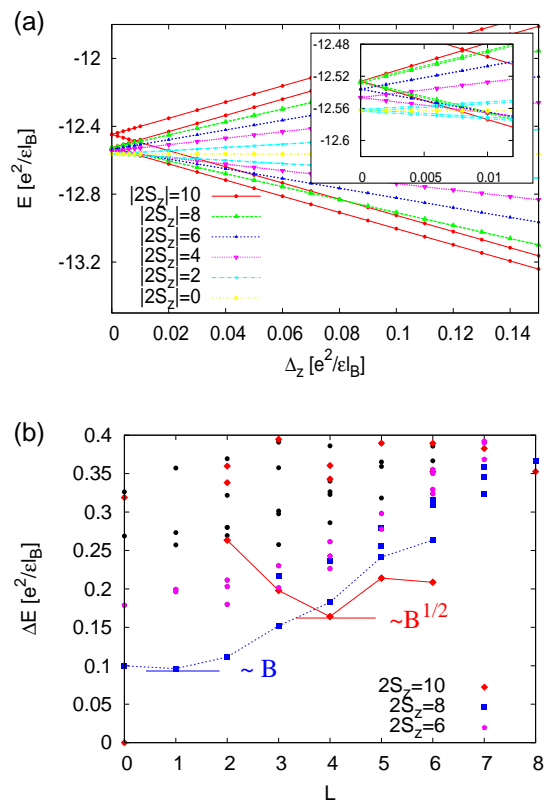}
\caption{\footnotesize{(Color online) {\sl (a)} Energy spectrum for $N=22$ electrons at a filling factor $\nu=1+1/3$, as a function
of $\Delta_Z$ in the different spin sectors $2S_z$, obtained from ED calculations of the Coulomb interaction on the sphere 
($N_B=15$) with implemented SU(2) symmetry. The inset shows a zoom on the region for small values of $\Delta_Z$.
{\sl (b)} Excitation spectrum at $\Delta_Z=0.05 e^2/\epsilon l_B$, as a function of the angular momentum $L$. The energy is
measured with respect to the ground state (at $L=0$). The spin-flip mode is found in the $2S_z=8$ sector (blue squares)
and scales linearly in $B$, whereas
the $2S_z=10$ sector reveals the usual magneto-roton branch (red diamonds), which scales as $\sqrt{B}$ with the $B$-field.
}}
\label{figSU2}
\end{figure}

Figure \ref{figSU2}(a) shows the energy spectrum for $N=22$ particles and $N_B=15$ flux quanta, in the different spin sectors,
obtained by ED of the SU(2) Coulomb interaction potential in the lowest LL. The spectrum is in qualitative agreement with
that obtained for the four-component system at $\nu=2+1/3$ (Fig. \ref{figSU4}) -- because the wave function (\ref{wave2}) is
not an eigenstate of the SU(2)-symmetric Coulomb potential, it does not describe the ground state at $\Delta_Z=0$, where
one obtains a three-fold degenerate state (with $2S_z=0,\pm 2$), but in a compressible sector
($L\neq 0$). As for the four-component case, a small symmetry-breaking Zeeman gap $\Delta_Z^1\simeq 0.01 e^2/\epsilon l_B$
suffices to stabilize a state with maximal spin polarization ($2S_z=10$ and $N_{\da}=6$), which has an overlap of $99\%$ with the wave
function (\ref{wave2}) \footnote{Notice that the subspace $2S_z=10$ is of dimension 6, instead of 1 for the subspace with
maximal spin polarization in the SU(4) calculation with $N=17$ electrons.}, and the lowest-lying
excited state in an intemediate range of the Zeeman gap, $\Delta_Z^1\leq \Delta_Z\leq \Delta_Z^2\simeq 0.08 e^2/\epsilon l_B$,
involves a spin flip as it is found in the spin sector $2S_z=8$. 

It has been argued that, for vanishing Zeeman splitting, the state at $\nu=1+1/3$
should be a spin-singlet 
composite-fermion (CF) state with reversed flux attachment \cite{jainHund}. Hund's rule, according to which the
system chooses a maximally polarized spin inside each energy level, would then predict an unpolarized state because $\nu=2/3$ 
corresponds to a completely filled lowest CF-LL \cite{jainHund}, but the same rule favors a completely polarized state if applied 
to the original electron coordinates. Our results indicate that already for a very small Zeeman splitting, a completely
polarized state is favored that satisfies the electronic instead
of the CF version of Hund's rule. Notice, however, that a direct numerical comparison between both states, CF spin singlet and
state (\ref{wave2}), is problematic in the spherical geometry because the spin-singlet state has a different flux-particle-number
relation, $N_B=(3/4)N-1$, than the polarized state (\ref{wave2}), $N_B=(3/4)N-3/2$. 
We find for $N=20$ and $N_B=14$ (results not shown) that the ground state is indeed a singlet at low Zeeman splittings,
but it is maximally polarized above
a value of $\Delta_Z/(e^2/\epsilon l_B)\sim 0.03$, which is on the same order of magnitude as $\Delta_Z^1$.

In order to gain further insight into the nature of the low-lying excitations, we have calculated the spectrum 
[Fig. \ref{figSU2}(b)] at an intermediate value of the Zeeman gap, $\Delta_Z=0.05 e^2/\epsilon l_B$, where spin-flip excitations
are expected to be relevant. The spectrum is now plotted as a function of the angular momentum in order to make apparent 
possible low-energy collective excitations of the incompressible state (\ref{wave2}). Within the charge sector with no change 
in the spin polarization, one observes in Fig. \ref{figSU2}(b) the usual magneto-roton branch (red diamonds) \cite{HR85}
which arises from gapped density-wave excitations \cite{GMP} and which is a prominent feature of Laughlin-type physics. 
However, another mode is apparent in Fig. \ref{figSU2}(b) that indicates the presence of collective excitations 
beyond the usual one-component Laughlin state and that is precisely a spin-flip excitation (blue squares). 
This mode, which is the lowest-energy excitation in the low-$L$ regime, is well separated from the high-energy part of the excitation
spectrum, such that it is likely to be a true collective mode. Notice that in the large-$L$ limit, the magneto-roton branch 
has a lower energy, and one may thus conjecture that the activation gap, i.e. the energy to create a well-separated 
quasiparticle-quasihole pair at large values of $L$, does not involve a spin-flip excitation, but is governed by 
one-component Laughlin physics. 

The relevance of collective spin-flip excitations in an intermediate Zeeman-gap range may eventually be tested experimentally in 
inelastic light-scattering experiments that are capable of probing collective excitations at finite wave vectors 
\cite{perspectives,eriksson}. Indeed, these experiments probe characteristic parts of the dispersion relation that 
show an enhanced density of states (such as at its minima and maxima). Because the spin-flip mode in Fig. \ref{figSU2}(b)
is almost flat at low angular momenta $L$ that correspond to small wave vectors, one may expect an enhanced peak in
such inelastic light-scattering measurements, at energies in the $0.1 e^2/\epsilon l_B$ range (roughly half of the energy
of the magneto-roton minimum, for the particular choice $\Delta_Z=0.05 e^2/\epsilon l_B$). As one may see in Fig. \ref{figSU2},
the spin-flip excitation scales linearly with the Zeeman gap, such that the associated peak is expected to scale linearly
with the magnetic field as well, whereas that of the usual magneto-roton would scale as $\sqrt{B}$ [see Fig. \ref{figSU2}(b)].
The observation of such a linear $B$-field dependence of the light-scattering peak would be clear
evidence for the relevance of spin-flip, beyond the properties of the Laughlin liquid, of the $\nu_G=1/3$ 
state in graphene.

In conclusion, we have shown, within ED calculations for a four- and a two-component system on the sphere,
how a FQHE can arise in graphene at $\nu_G=1/3$ even at very small values of 
a spin-valley symmetry-breaking Zeeman field. Although the leading energy scale is given by the SU(4)-invariant Coulomb
interaction, a small Zeeman gap $\Delta_Z/(e^2/\epsilon l_B)\sim 0.01$ is sufficient to fully polarize the electronic 
spin and thus to stabilize the state (\ref{wave1}) which we have identified as being responsible for the observed graphene
FQHE \cite{du09,bolotin09}. 
In spite of its reminiscence with the Laughlin state, novel collective excitations that are inherent to the four-component
character of graphene determine the low-energy spectrum at intermediate values of the Zeeman gap, $\Delta_Z^1\leq \Delta_Z\leq
\Delta_Z^2$, that correspond to the
experimental situation in which the FQHE has been observed. In order to gain further insight into the nature of these
spin-flip excitations, which may be visible in inelastic light-scattering experiments, we have performed ED calculations 
in an analogous two-component quantum Hall system at a filling factor $\nu=1+1/3$ that corresponds to a completely filled
spin-$\ua$ and a one-third filled spin-$\da$ branch. The spin-flip excitation is well separated
from the high-energy part of the energy spectrum thus indicating its collective nature, in addition to the usual magneto-roton
branch that determines the low-energy spectrum in the large angular-momentum regime.



This work was supported by Agence Nationale de la Recherche under Grant No. ANR-JCJC-0003-01 and by grants
from R\'egion Ile-de-France. Z.P. was furthermore supported by the Serbian Ministry of
Science under Grant No. 141035. We acknowledge fruitful discussions with Ch. Glattli and M. Milovanovi\'c.



\begin{thebibliography}{99}


\bibitem{du09}
X. Du, I. Skachko, F. Duerr, A. Luican, and E. Y. Andrei,
Nature {\bf 462}, 192 (2009).

\bibitem{bolotin09}
K. I. Bolotin, F. Ghahari, M. D. Shulman, H. L. Stormer, and P. Kim,
Nature {\bf 462}, 196 (2009).


\bibitem{peres06}
N. M. R. Peres, F. Guinea, and A. H. Castro Neto, Phys. Rev. B {\bf 73}, 125411 (2006).

\bibitem{GMD}
M. O. Goerbig, R. Moessner, and B. Dou\c cot,
Phys. Rev. B {\bf 74}, 161407 (2006).

\bibitem{AC06}
V. M. Apalkov and T. Chakraborty, Phys. Rev. Lett. {\bf 97}, 126801 (2006);
C. T\"oke, P. E. Lammert, V. H. Crespi, and J. K. Jain, Phys.
Rev. B {\bf 74}, 235417 (2006).

\bibitem{YDSMD06}
K. Yang, S. Das Sarma, and A. H. MacDonald, Phys. Rev. B {\bf 74}, 075423 (2006).

\bibitem{khvesh07}
D. V. Khveshchenko, Phys. Rev. B {\bf 75}, 153405 (2007).

\bibitem{toke07} C. T\"oke and J. K. Jain, Phys. Rev. B {\bf 75}, 245440 (2007);
M. O. Goerbig and N. Regnault, Phys. Rev. B {\bf 75}, 241405 (2007);
for a review on the theory of the FQHE in graphene, see 
Z. Papi\'c, M. O. Goerbig, and N. Regnault, Solid State Comm. {\bf 149}, 1056 (2009).

\bibitem{TSG}D.\ C.\ Tsui, H.\ Stormer, and A.\ C.\ Gossard, Phys.\ Rev.\ Lett.\ {\bf 48}, 1559 (1982).

\bibitem{laughlin}R.\ B.\ Laughlin, Phys.\ Rev.\ Lett.\ {\bf 50}, 1395 (1983).


\bibitem{zhang2}Y. Zhang, Z. Jiang, J. P. Small, M. S. Purewal, Y.-W. Tan, M. Fazlollahi, 
J. D. Chudow, J. A. Jaszczak, H. L. Stormer, and P. Kim, Phys. Rev. Lett, {\bf 96}, 136806 (2006).


\bibitem{eriksson}
M. A. Eriksson, A. Pinczuk, B. S. Dennis, S. H. Simon,
L. N. Pfeiffer, and K. W. West, Phys. Rev. Lett. {\bf 82}, 2163 (1999);
Y. Gallais, T. H. Kirschenmann, I. Dujovne, C. F. Hirjibehedin, A. Pinczuk, B. S. Dennis, L. N. Pfeiffer, and K. W. West,
Phys. Rev. Lett. {\bf 97}, 036804 (2006).


\bibitem{perspectives}For a review, see A. Pinczuk in
{\sl Perspectives in Quantum Hall Effects}, edited by S.\ Das\ Sarma and 
A.\ Pinczuk (John Wiley, New York, 1997).



\bibitem{AF}
J. Alicea and M. P. A. Fisher,
Phys. Rev. B {\bf 74}, 075422 (2006).

\bibitem{DGRG}
R. de Gail, N. Regnault, and M. O. Goerbig, Phys. Rev. B {\bf 77}, 165310 (2008). 

\bibitem{haldane}F. D. M. Haldane, Phys. Rev. Lett. {\bf 51}, 605 (1983).

\bibitem{GGV} J. Gonz\'alez, F. Guinea, and M. A. H. Vozmediano, Phys. Rev. B {\bf 59}, R2474 (1999).

\bibitem{RFG}
K. Shizuya, Phys. Rev. B {\bf 75}, 245417 (2007);
R. Rold\'an, J.-N. Fuchs, and M. O. Goerbig, Phys. Rev. B {\bf 80}, 085408 (2009).

\bibitem{goerbigRev} For a recent review, see M. O. Goerbig, arXiv:1004.3396, and references therein.

\bibitem{jainHund}X. G. Wu, G. Dev, and J. K. Jain, Phys. Rev. Lett. {\bf 71}, 153 (1993); J. K. Jain and X. G. Wu,
Phys. Rev. B {\bf 49}, 5085 (1994).

\bibitem{HR85} F. D. M. Haldane et E. H. Rezayi, Phys. Rev. Lett. {\bf 54},
237 (1985); G. Fano, F. Ortolani et E. Colombo, Phys. Rev. B {\bf 34},
2670 (1986).

\bibitem{GMP}S.\ M.\ Girvin, A.\ H.\ MacDonald et P.\ M.\ Platzman, 
Phys.\ Rev.\ B\ {\bf 33}, 2481 (1986).


\end{thebibliography}
\end{document}